\documentclass[12pt]{article}
 \usepackage{epsfig}
  \usepackage{wrapfig}
   \usepackage{graphicx}
\begin{document}

\begin{titlepage}
\begin{center}

{\Large\bf  A possible resolution of the strange quark
polarization puzzle ?}

\end{center}
\vskip 2cm
\begin{center}
{\bf Elliot Leader}\\
{\it Imperial College London\\ Prince Consort Road, London SW7
2BW, England }
\vskip 0.5cm
{\bf Alexander V. Sidorov}\\
{\it Bogoliubov Theoretical Laboratory\\
Joint Institute for Nuclear Research, 141980 Dubna, Russia }
\vskip 0.5cm
{\bf Dimiter B. Stamenov \\
{\it Institute for Nuclear Research and Nuclear Energy\\
Bulgarian Academy of Sciences\\
Blvd. Tsarigradsko Chaussee 72, Sofia 1784, Bulgaria }}
\end{center}

\vskip 0.3cm
\begin{abstract}
\hskip -5mm

The strange quark polarization puzzle, i.e. the contradiction
between the negative polarized strange quark density obtained from
analyses of inclusive DIS data and the positive values obtained
from combined analyses of inclusive and semi-inclusive SIDIS data
using de Florian et. al. (DSS) fragmentation functions, is
discussed. To this end the results of a new combined NLO QCD
analysis of the polarized inclusive and semi-inclusive DIS data,
using the Hirai et. al. (HKNS) fragmentation functions, are
presented. It is demonstrated that the polarized strange quark
density is very sensitive to the kaon fragmentation functions, and
if the set of HKNS fragmentation functions is used, the polarized
strange quark density obtained from the \emph{combined} analysis
turns out to be \emph{negative} and well consistent with values
obtained from the pure DIS analyses.


\end{abstract}

\end{titlepage}

\newpage
\setcounter{page}{1}

In the absence of neutrino reactions on a polarized target, the
inclusive polarized deep inelastic lepton-hadron reactions
determine only the sum of quark and anti-quark polarized parton
density functions (PDFs), $\Delta q(x) + \Delta \bar{q}(x)$, and
provide no information at all about the individual polarized
anti-quark densities. All analyses of the polarized
\emph{inclusive} (DIS) data have produced results for the
polarized strange quark density function, $\Delta s(x) + \Delta
\bar{s}(x)$, which are significantly \emph{negative} for all
values of $x$ (for more recent analyses see \cite
{{groups},{LSS07}}). One way to determine polarized quark and
anti-quark densities separately is to use the data on polarized
\emph{semi-inclusive} reactions (SIDIS) like $l+p\rightarrow l + h
+ X $, where $h$ is a detected hadron. In the past few years more
data on polarized SIDIS processes have become available and have
led to assertions that $\Delta s(x) + \Delta \bar{s}(x)$ is
\emph{positive} for most of the range of measured $x$. In the
following we discuss possible resolutions to this puzzling state
of affairs.

It should be noted that in the study of the SIDIS data it is usual
to simplify the analysis by taking $\Delta s(x) = \Delta
\bar{s}(x)$. On the other hand, it has been suggested, on the
basis of theoretical models, that this equality is badly broken
\cite{Brodsky}, and that this could be the cause of the conflict.
However, it is crucial to realize that: (i) the DIS result for
$\Delta s(x) + \Delta \bar{s}(x)$ is independent of the
relationship between $\Delta s(x)$ and $ \Delta \bar{s}(x)$, and
(ii) that the  COMPASS estimate \cite{COMPASS_dels} of the
difference $\Delta s(x) - \Delta \bar{s}(x)$ is much smaller than
the theoretical model estimates and thus cannot be the cause of a
serious error in the extraction of $\Delta s(x) + \Delta
\bar{s}(x)$ from a combined analysis of the DIS and SIDIS data.

The key to resolving the puzzle lies, we believe, in the
properties of the fragmentation functions (FFs) needed in the
theoretical expressions for the measured SIDIS cross-sections and
asymmetries, which involve convolutions of either unpolarized or
polarized PDFs with the FFs. There are three modern versions of
the FFs in the literature, Hirai et al. (HKNS) \cite{HKNS}, de
Florian et al. (DSS) \cite{DSS} and Albino et. al. (AKK)
\cite{AKK}, sometimes differing significantly from each other.
They are based mainly on semi-inclusive $e^+ \, e^-$ annihilation
data (HKNS), $e^+ \, e^-$ annihilation and RHIC data on reactions
like $pp\rightarrow \pi \, \textrm{or} \, K + X$ (AKK), and a
global analysis (DSS) of the data on semi-inclusive $e^+ \, e^-$
annihilation, the proton-proton collisions at RHIC and unpolarized
SIDIS processes.

The early claim by the HERMES Collaboration \cite{HERMESpos_dels}
that the polarized SIDIS data implied marginally positive $\Delta
s(x)+ \Delta \bar{s}(x)$ in the measured $x$ range [0.023-0.3] was
based on a LO analysis of the data. In 2008, de Florian, Sassot,
Stratmann and Vogelsang (DSSV) carried out a combined NLO QCD
analysis \cite{DSSV} of polarized DIS, SIDIS and RHIC data using
the DSS fragmentation functions and effectively confirmed the LO
result. More precisely, using the assumption $\Delta s(x)=\Delta
\bar{s}(x)$ they obtained a sign-changing solution for $\Delta
s(x) + \Delta \bar{s}(x)$, negative for $x < 0.03$ and positive in
the region $x > 0.03$. Later we repeated this analysis
\cite{LSS10}, using polarized DIS and SIDIS data and found
substantial agreement with DSSV. We confirmed the sign changing
behavior of $\Delta \bar{s}(x)$,  though our $\Delta \bar{s}(x)$
is less negative at $x < 0.03$ and less positive for large x and
compatible with zero within the errors. Note that the polarized pp
data from RHIC are not important for the determination of the
polarized quark and anti-quark densities; they constrain mainly
the gluon polarization.

After convincing ourselves that the puzzle could not be resolved
by taking $\Delta s(x) \neq \Delta \bar{s}(x)$  we began to try to
test whether the problem lay in the properties of  the  FFs. Now
the largest disagreements between the various sets of FFs in the
literature occur in the kaon production sector. To this end we
first carried out a combined  NLO QCD analysis
\cite{only_pionSIDIS} of the polarized world DIS data \cite{DIS}
and \emph{just the pion} SIDIS data
\cite{{COMPASS_dels},{pionSIDIS}}, using the DSS FFs. Note that in
this case only the sum $x(\Delta s +\Delta \bar{s})(x, Q^2)$ can
be determined from the data because of the reasonable assumption
$D_s^{\pi}=D_{\bar{s}}^{\pi}$ used for all the sets of the
fragmentation functions. The result for $x(\Delta s +\Delta
\bar{s})/2$ is illustrated in Fig. 1 and compared to those
obtained from the LSS'06 DIS analysis \cite{LSS07} (red curve) and
the combined LSS'10 fit to the DIS and SIDIS data \cite{LSS10}
(black curve). As seen from Fig. 1, in the presence
 of only the $A_{1N}^{\pi}$ data, $x(\Delta s(x) +\Delta
\bar{s}(x))/2$ (blue curve) is still \emph{negative} in the
measured $x$ region as in the  analyses of the purely inclusive
data.
\begin{figure}[h]
\begin{center}
\resizebox{0.45\hsize}{0.25\vsize}{\includegraphics{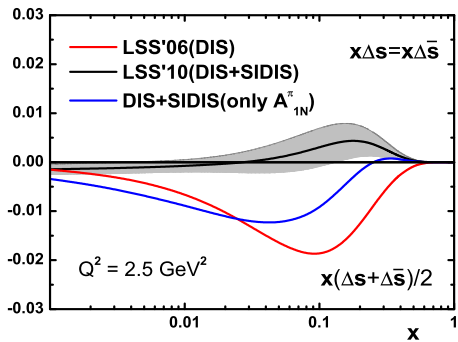}}
\caption{Comparison between polarized strange quark densities
obtained from different kinds of NLO QCD analyses (see the text).}
\end{center}
\end{figure}

This definitely seemed to point towards the kaon FFs as the source
of the conflict. Note also that it had  already been pointed out
by COMPASS Collaboration (2nd and 3rd refs in \cite{pionSIDIS})
that in the LO QCD approximation the value of the first moment of
$\Delta s(x)$ in the measured range of $x$ is very sensitive to
the assumed value of the ratio of the $\bar{s}$-quark to $u$-quark
fragmentation functions into positive kaons. Therefore, we carried
out a new combined NLO QCD analysis of the polarized DIS and
\emph{all} the SIDIS data
\cite{{COMPASS_dels},{pionSIDIS},{hadronSIDIS}} using the HKNS set
of FFs \cite{HKNS}, which  differ significantly from the DSS ones
in the kaon sector, especially for the transition
$\bar{s}\rightarrow K^+$, as shown in Fig 2 \cite{note}. In Fig. 2
two error bands for the HKNS FFs are presented. The narrow one
corresponds to $\Delta \chi^2 =1$ while the wide corridor
corresponds to $\Delta \chi^2 =19.2$. The latter value corresponds
to 17 parameters fit in the MINUIT-procedure when only the
statistical errors are taken into account. However, the authors of
\cite{HKNS} apply this procedure for the statistical and
systematic errors added in quadrature which definitely
overestimates the uncertainties.
\begin{figure}[ht]
\begin{center}
\resizebox{0.80\hsize}{0.43\vsize}{\includegraphics{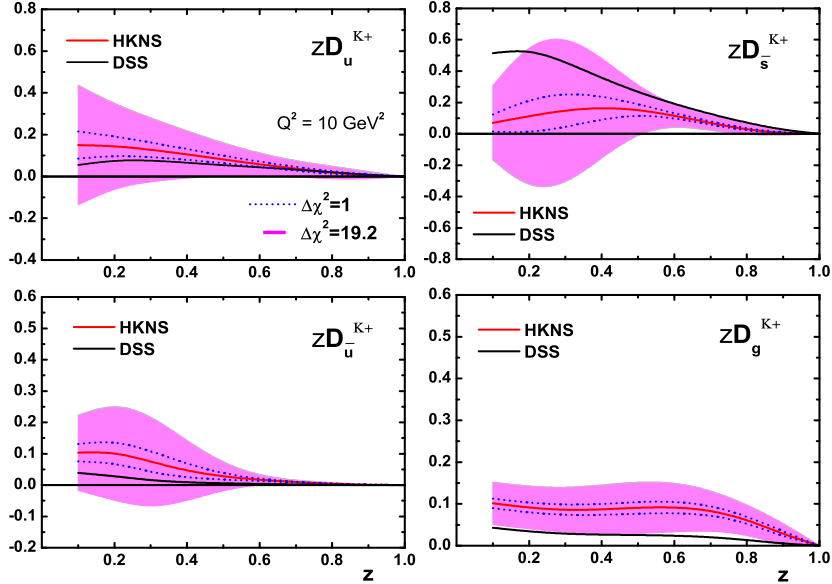}}
\caption{Comparison between NLO HKNS and DSS kaon FFs at
$Q^2=10~GeV^2$.}
\end{center}
\end{figure}

The method used is the same as in our previous analysis
\cite{LSS10} of the same set of data when the DSS FFs were used.
Note  that the present SIDIS data are not precise enough to
determine separately $\Delta s(x)$ and $\Delta \bar{s}(x)$. So, as
in our previous analysis the assumption $\Delta s(x)=\Delta
\bar{s}(x)$ was used. A good description of the SIDIS data
($\chi^2_{NrP}$=0.92) is achieved using the HKNS FFs (NrP is the
number of corresponding experimental points). The quality of the
fit to the data is demonstrated in Fig. 3 (black curves) for some
of the SIDIS asymmetries obtained by the HERMES and COMPASS
Collaborations. The new curves are compared to our previous
theoretical curves (red ones) obtained from the best fit to the
data using the DSS FFs ($\chi^2_{NrP}$=0.87). As seen from Fig. 3
the results from both the fits are very close to each other and
for some of the asymmetries the curves are almost identical.
\begin{figure}[ht]
\begin{center}
\resizebox{0.97\hsize}{0.40\vsize}{\includegraphics{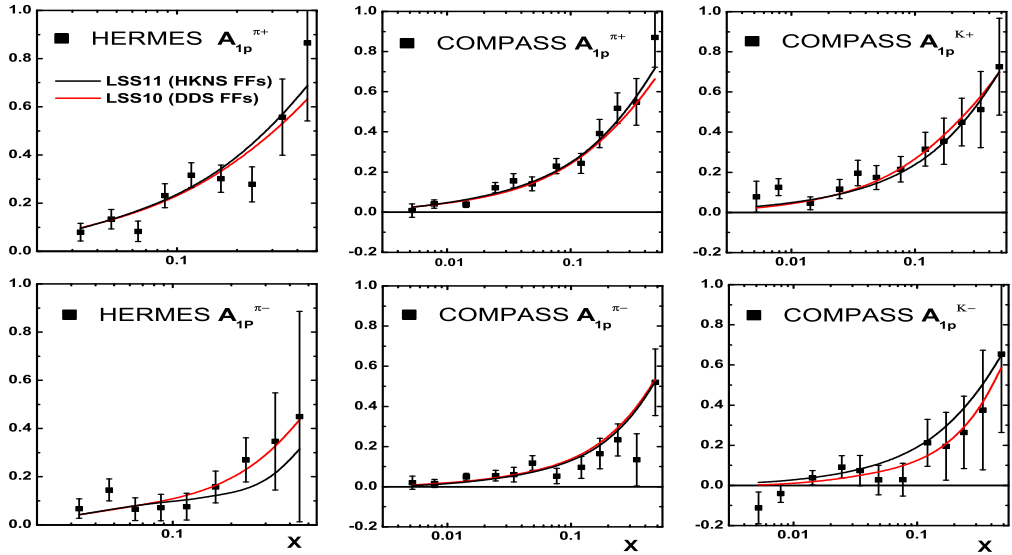}}
\caption{Comparison of our NLO LSS'11 (black curves) and LSS'10
(red curves) results for the SIDIS asymmetries with the data at
measured $x$ and $Q^2$. }
\end{center}
\end{figure}
\begin{figure}[t]
\begin{center}
\resizebox{0.75\hsize}{0.43\vsize}{\includegraphics{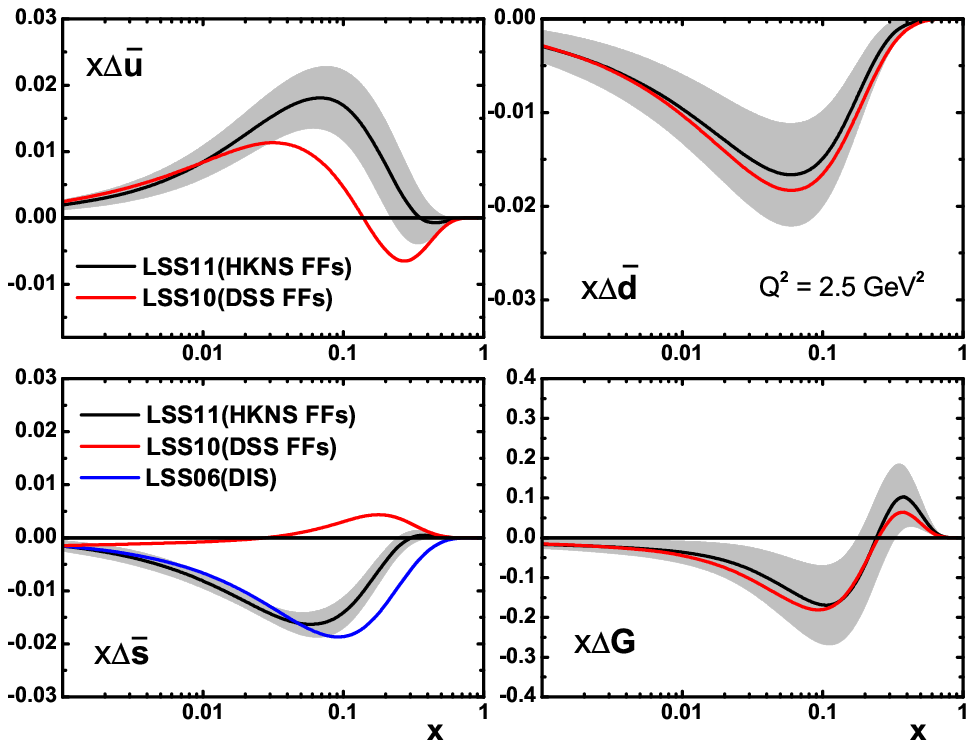}}
\caption{Comparison between NLO LSS'11(HKNS FFs) and LSS'10(DSS
FFs) sea quark and gluon polarized PDFs at $Q^2=2.5~GeV^2$. The
blue curve corresponds to $x(\Delta s(x)+\Delta \bar{s}(x))/2$
obtained from the pure DIS analysis \cite{LSS07}.  }
\end{center}
\end{figure}

Let us discuss the impact of the HKNS fragmentation functions on
the polarized sea-quark densities. It is known that the present
SIDIS data do not influence the gluon polarization. It is mainly
determined from inclusive DIS and semi-inclusive $pp$ RHIC data.
The new values of the sea quark and gluon polarized densities
(black curves) are presented in Fig. 4 together with their error
bands and compared to those obtained using the DSS FFs (LSS'10).
As seen from Fig. 4 the changes in the polarized sea quark
densities are as follows: negligible for $x\Delta \bar{d}(x)$,
visible for $x\Delta \bar{u}(x)$ at $x > 0.03$ and dramatic for
$x\Delta \bar{s}(x)$, although the central values of the first
moments of $\Delta \bar{s}$(DSS) and $\Delta \bar{s}$(HKNS) are
very close to each other ($-0.052 \pm 0.016$ and $-0.048 \pm
0.012$ at $Q^2=1~GeV^2$ for DSS and HKNS FFs, respectively) and
coincide within the errors. In Fig. 4 our LSS'06 result
\cite{LSS07} for $x(\Delta s(x)+\Delta \bar{s}(x))/2$ (blue curve)
obtained from the NLO QCD analysis of the world inclusive DIS data
is presented too. We find now that if the HKNS FFs are used,
$\Delta \bar{s}(x)$ is \emph{negative} and well consistent with
$(\Delta s(x) + \Delta \bar{s}(x))/2$ obtained from the pure DIS
analyses \cite{{groups},{LSS07}}.\\

In conclusion, we have found that in the presence of
semi-inclusive DIS data the strange quark density is very
sensitive to the choice of the FFs. We have also demonstrated that
the strange quark polarization puzzle can be resolved by using the
HKNS set of fragmentation functions rather than the DSS ones.
Finally, we like to stress we do not claim to have presented a
unique resolution to the strange polarization puzzle. Our analysis
illustrates only how badly we need to have a more reliable
determination of FFs in order to extract correctly the polarized
sea quark densities. To this end precise \emph{unpolarized} SIDIS
cross-section data are very important.

\vskip 0.6cm {\it Acknowledgments:} We thank S. Kumano and M.
Hirai for the Fortran code of the HKNS fragmentation functions and
some additional explanations. One of us (D. S) is also thankful to
R. Sassot for  useful discussions. This research was supported by
the JINR-Bulgaria Collaborative Grant, by the RFBR Grants (No
09-02-01149 and 11-01-00182) and by the Bulgarian National Science
Fund under Contract 02-288/2008.


\begin{thebibliography}{99}

\bibitem{groups}
M. Hirai, S. Kumano, and N. Saito (Asymmetry Analysis
Collaboration), Phys. Rev. D {\bf 74}, 014015 (2006); V.Y.
Alexakhin {\it et al.} (COMPASS Collaboration), Phys. Lett. B {\bf
647}, 8 (2007); J. Blumlein and H. B\"{o}ttcher, Nucl. Phys. {\bf
B841}, 205 (2010).

\bibitem{LSS07}
E. Leader, A.V. Sidorov, and D.B. Stamenov, Phys. Rev. D {\bf 75},
074027 (2007).

\bibitem{Brodsky}
S. J. Brodsky and Bo-Qiang Ma, Phys. Lett. B {\bf 381}, 317
(1996); Y. Ding, Rong-Guang Xu, and Bo-Qiang Ma, Phys. Rev. D {\bf
71}, 094014 (2005).

\bibitem{COMPASS_dels}
M.G. Alekseev {\it et al.} (COMPASS Collaboration), Phys. Lett. B
{\bf 693}, 227 (2010).

\bibitem{HKNS}
M. Hirai, S. Kumano, T.-H. Nagai, and K. Sudoh, Phys. Rev. D {\bf
75}, 094009 (2007).

\bibitem{DSS}
D. de Florian, R. Sassot, and M. Stratmann, Phys. Rev. D {\bf 75},
114010 (2007); Phys. Rev. D {\bf 76}, 074033 (2007).

\bibitem{AKK}
S. Albino, B. A. Kniehl, and G. Kramer, Nucl. Phys. B {\bf 803},
42 (2008).

\bibitem{HERMESpos_dels}
U. Stosslein, Acta Phys. Polonica B {\bf 33}, 2813 (2002); A.
Airapetian et al. (HERMES Collaboration), Phys. Rev. Lett. {\bf
92}, 012005 (2004).

\bibitem{DSSV}
D. de Florian, R. Sassot, M. Stratmann, and W. Vogelsang, Phys.
Rev. {\bf D 80}, 034030 (2009).

\bibitem{LSS10}
E. Leader, A.V. Sidorov, and D.B. Stamenov, Phys. Rev. D {\bf 82},
114018 (2010).

\bibitem{only_pionSIDIS}
E. Leader, A.V. Sidorov, and D.B. Stamenov, arXiv:1012.5033
[hep-ph] (to be published in Eur. Phys. J. ST).

\bibitem{DIS}
J. Ashman {\it et al.} (EMC Collaboration), Phys. Lett. B {\bf
206}, 364 (1988); Nucl. Phys. {\bf B328}, 1 (1989); P.L. Anthony
{\it et al.} (SLAC E142 Collaboration), Phys. Rev. D {\bf 54},
6620 (1996); K. Abe {\it et al.} (SLAC E143 Collaboration), Phys.
Rev. D {\bf 58}, 112003 (1998); K. Abe {\it et al.} (SLAC/E154
Collaboration), Phys. Rev. Lett. {\bf 79}, 26 (1997); P.L. Anthony
{\it et al.} (SLAC E155 Collaboration), Phys. Lett. B {\bf 493},
19 (2000); Phys. Lett. B {\bf 463}, 339 (1999); A. Airapetian {\it
et al.} (HERMES Collaboration), Phys. Rev. D {\bf 71}, 012003
(2005); X. Zheng {\it et al.} (JLab/Hall A Collaboration), Phys.
Rev. Lett. {\bf 92}, 012004 (2004); Phys. Rev. C {\bf 70}, 065207
(2004); K.V. Dharmwardane {\it et al.} (CLAS Collaboration), Phys.
Lett. B {\bf 641}, 11 (2006); B. Adeva {\it et al.} (SMC
Collaboration), Phys. Rev. D {\bf 58}, 112001 (1998); V.Yu.
Alexakhin {\it et al.} (COMPASS Collaboration), Phys. Lett. B {\bf
647}, 8 (2007); M.G. Alekseev {\it et al.} (COMPASS
Collaboration), Phys. Lett. B {\bf 690}, 466 (2010).

\bibitem{pionSIDIS}
A. Airapetian {\it et al.} (HERMES Collaboration), Phys. Rev. D
{\bf 71}, 012003 (2005); M.G. Alekseev {\it et al.} (COMPASS
Collaboration), Phys. Lett. B {\bf 680}, 217 (2009); Phys. Lett. B
{\bf 690}, 466 (2010).

\bibitem{hadronSIDIS}
B. Adeva {\it et al.} (SMC Collaboration), Phys. Lett. B {\bf
420}, 180 (1998); M.G. Alekseev {\it et al.} (COMPASS
Collaboration), Phys. Lett. B {\bf 660}, 458 (2008).

\bibitem{note}
The significant difference in the kaon sector between the DSS FFs
and the other sets of FFs, including HKNS, is due to the
unpublished HERMES'05 data on the unpolarized hadron
multiplicities used only in the DSS analysis. We consider that
until the final HERMES and COMPASS data on the hadron
multiplicities are presented, the question of a more reliable
determination of the fragmentation functions in the $Q^2$ range of
the present unpolarized SIDIS processes is open.

\end{thebibliography}
\end{document}